\documentclass[numbers,webpdf,imaiai]{bakwashe}%
\usepackage{hyperref}
\hypersetup{
    colorlinks=true,
    linkcolor=blue,
    filecolor=blue,      
    urlcolor=blue,
    }

\graphicspath{{Fig/}}

\theoremstyle{thmstyletwo}%
\numberwithin{equation}{section}

\begin{document}

\DOI{DOI HERE}
\copyrightyear{2021}
\vol{00}
\pubyear{2021}
\access{Advance Access Publication Date: Day Month Year}
\appnotes{Paper}
\firstpage{1}


\title[Are Physical Theories Incommensurable?]{Are Physical Theories Incommensurable?\footnote{By physical theories, I mean the theories that fall under the domain of physics.}}

\author{Kharanshu Solanki$^\dagger$
\address{\orgdiv{School of Arts and Sciences}, \orgname{Ahmedabad University}, \orgaddress{\street{Ahmedabad}, \postcode{India}, \state{380009}}}}

\authormark{Kharanshu Solanki}

\corresp[$\dagger$]{\href{email:email-id.com}{kharanshu.s@ahduni.edu.in}}

\received{Date}{0}{Year}


\abstract{\noindent This paper examines the incommensurability thesis - one of the most important and controversial ideas to emerge from the simultaneous work of Kuhn and Feyerabend. In the first half, I discuss three aspects of incommensurability - \emph{methodological incommensurability} (the view that each paradigm supplies different standards of evaluation), \emph{observational incommensurability} (the view that the theories we accept alter how we see the world), and finally, \emph{semantic incommensurability} (which claims that as paradigms change, the very meanings of central theoretical terms also change). In the latter half, I tackle the general arguments of physicists against incommensurability by primarily considering the so called unifying \emph{cube of physics}. I show that the cube of physics does not get rid of incommensurability, but rather favours it.}


\maketitle
\vspace{2em}
\section{Introduction}
In 1962, Kuhn gave us the idea that scientists adopt \emph{paradigms} (which are basically whole ways of doing science). Kuhn's model \cite{Kuhn1962} is that we have normal science governed by a single paradigm, where \emph{anomalies} build up, eventually leading to a \emph{crisis}, where scientists lose confidence in the paradigm. This leads to a \emph{revolution}, where one paradigm is replaced by another and normal science resumes. According to Kuhn, competing paradigms are \emph{incommensurable}, which basically means that they have \emph{no common measure for comparison}. This is one notion of incommensurability that we get from Kuhn. However, the concept of incommensurability goes beyond the absence of comparison bases. Some of the most interesting aspects of incommensurability were put forth by Paul Feyerabend (also in 1962). In fact, Feyerabend's notions of incommensurability \cite{feyerabend_1981} are not completely independent of Kuhn's notions, because he referred to Kuhn's then unpublished work while developing his ideas.

\section{Three Aspects of Incommensurability}

\noindent The concept of incommensurability (as developed by Kuhn and Feyerabend) can be broken down into three key ideas: 
\begin{itemize}
    \item Methodological incommensurability.
    \item Observational incommensurability.
    \item Semantic incommensurability.
\end{itemize}

\noindent All these aspects are quite important and convey the essence of incommensurability. It therefore becomes necessary that we understand these three aspects before arguing for or against the role of incommensurability in physics\footnote{I will concern myself with the relevance of incommensurability only to physics.} (or any other science for that matter). 

\subsection{Methodological Incommensurability}

The methodological aspect of incommensurability is exactly the one that I already mentioned. This is the claim that since each paradigm supplies its own standards of evaluation, there are no independent standards by which one can evaluate competing paradigms \cite{Hoyningen-Huene1990-HOYKCO}. In other words, one cannot step outside the realms of the competing paradigms to find an \emph{objective} standard of evaluation. Methodological incommensurability can be well understood by considering the case of gravitational theories before Einstein, i.e., Newton's and Descartes' theories of gravity.\\

\noindent One major problem with Newton's theory \cite{https://doi.org/10.48550/arxiv.1807.01791} was that it assumed \emph{non-locality}\footnote{Non-locality is the idea that information can instantaneously propagate from one point in space to another. This is a violation of the all-important postulates of relativity.} rather than explaining \emph{how} the gravitational force propagated. This \emph{action-at-a-distance} approach to gravity seemed quite mysterious to the physicists of Newton's time because they adopted the \emph{mechanistic approach to science}, where causes were supposed to be mechanistically linked to their effects.\\

\noindent A competing theory of gravity was offered by Descartes \cite{doi:10.1080/00033795900200038}, which claimed that all of space is filled with very fine matter, implying the non-existence of a \emph{truly empty space}. This matter tends to generate circular motion, which produces large \emph{vortices} in space. Descartes argued that the centrifugal force causes the fine matter to settle at the outer edges of these vortices, and subsequently, the inward pressure due to the fine matter pushes the heavier matter (like planets and stars) inwards. This pressure accounted for the effect of gravity.\\

\noindent Descartes' theory of vortices  is superior to that of Newton's in terms of providing a mechanistic explanation. On the other hand, Newton's theory produced \emph{better predictions}, which lead to its triumph over Descartes' theory. The interesting aspect of all this is that once Newton's theory was accepted into the scientific community, its lack of mechanistic explanations was no longer considered a problem. In fact, scientists altogether dropped the requirement that science must provide mechanistic explanations. Eventually, other scientists would have considered Newton's theory as an \emph{exemplar}\footnote{For instance, the theory of the electrostatic force developed by Coulomb is strikingly similar to Newton's theory of gravitation - both in terms of the equations, and in terms of the lack of a mechanistic explanation.} in the Kuhnian sense of the word.\\

\noindent The idea that we get from this historical account is that when the basis for evaluation of theories is changed, the problems of already existing theories tend to convert into virtues for any prospective theories. We see that due to changing methodological standards, Newton's theory became a paradigm - an exemplar against which one could judge other theories. The problems with Newton's theory were not solved by solving these problems but by altogether discarding the methodologies that produced those problems in the first place. Newton was more predictively accurate over Descartes while Descartes provided a mechanistic explanation. So, what is more important? Predictive accuracy or a mechanistic explanation? These are competing standards. One triumphed over the other, but did that have to be the case? It seems that scientists could have adopted for a different standard of evaluation and it is not clear that this choice would have been irrational or not.\\

\noindent In his later work, Kuhn allowed for certain standards to be widely accepted across paradigms. These include things like accuracy, consistency, scope, simplicity and fruitfulness for further research. But these virtues are imprecise and can be weighted in different ways. For instance, there is no algorithm to decide whether a theory is simple. Therefore, these virtues do not solve the problem of methodological incommensurability.

\subsection{Observational Incommensurability}

\noindent A rather more radical aspect of incommensurability is the observational aspect. This is more commonly known as the \emph{theory-ladenness of observation} \cite{Brewer2001-BRETTO}. The basic claim here is that the theories we accept alter the way we perceive the world. Standardly, we think of observations as providing a neutral arbiter between theories. For instance, if theory $A$ predicts $X$ and theory $B$ predicts $Y$, then it is straightforward to see which theory supports the observations. But according to Kuhn, the observations one makes are dependent on the theory that one accepts, which rejects the notion of the \emph{neutrality} of observations.\\

\noindent Norwood Russell Hanson gave a good example to understand this \cite{Hanson1958-HANPOD-2}. Suppose that Tycho Brahe and Johannes Kepler are watching the sun rise. Now, Brahe accepted the theory that the earth is the fixed center of the universe, while Kepler accepted that the earth orbits the sun. So, while observing the sun rise, Brahe would note down that the sun rises above a stationary horizon, while Kepler would note down that a moving horizon dips below the sun\footnote{One may argue that both these observations are correct, and that the only reason these observations differ is because the observers are in altogether different frames of reference (the relativistic argument), but note that Brahe and Kepler could be standing right next to each other (in the same frame of reference) and still make different observations.}. As Hanson puts it,

\begin{center}
   \emph{ There is more to seeing than meets the eyeball.} \cite{Hanson1958-HANPOD-2}
\end{center}

\noindent Kuhn provides another example for this in the context of particle physics:

\begin{center}
\noindent \emph{ Looking at a bubble-chamber photograph, the student\\ sees confused and broken lines, the physicist\\ a record of familiar subnuclear events.} \cite{Kuhn1962}
\end{center}

\noindent These examples, and more recent studies show that observations are indeed theory-laden. Hence, even if different paradigms had shared standards of evaluation, one would not be able to use evidence and logic alone to decide which one is better. Since scientists who adopt different paradigms see the world differently, the evidence is literally different for different paradigms. This implies that the decision to adopt to one of the competing paradigms is not a rational decision. In a nutshell, the data cannot settle the matter because the theories partially decide the data.\\

\noindent An obvious objection to this radical argument is that in order to say that there is no theory-neutral observation, one must show that this is true explicitly for all possible observations. These observations go well beyond what Kuhn and others have to offer. Other objections claim that the theory-ladenness of observations can be eliminated by careful reconsideration of the data. A seemingly good objection to observational incommensurability comes from Bogen and Woodward \cite{bogen1988saving}, who claim that theories are not tested against raw data, but against \emph{phenomena} which are unobservable and inferred from the data. Since phenomena are unobservable, they are not theory-laden, and hence one might get rid of observational incommensurability. 

\subsection{Semantic Incommensurability}

\noindent The semantic aspect of incommensurability was popularized primarily by Feyerabend. The basic idea of semantic incommensurability is that the very meaning of the central terms in competing theories or paradigms are different (this is also called the problem of \emph{meaning invariance}), or as Feyerabend puts it, there is \emph{deductive disjointedness} between theories.\\

\noindent A great way to understand semantic incommensurability is by considering the development of physical ideas from Newton to Einstein. There is a clear disjointedness when it comes to the meaning of fundamental terms such as mass, energy, space, time, etc. in Newtonian mechanics as compared to Einstein's relativistic theories. In Newtonian mechanics, space and time are independent of each other and referred to as \emph{absolute space and time}. For Newton, space is flat and homogeneous, with a Euclidean structure imposed on it. Further, it is assumed that \emph{space and time exist of their own accord}, independent of mass and energy (matter). Einstein, on the other hand rejects the notion of absolute space and time, and puts forth the idea that space, time, mass and energy are all correlated \cite{Einstein:1905ve}. To quote John Wheeler, 

\begin{center}
   \emph{Spacetime tells matter how to move; matter tells space-time how to curve}. \cite{thorne2000gravitation}
\end{center}

\noindent This is reflected in Einstein's famous field equations for gravitation \cite{Einstein:1915ca},

\begin{equation}
    R_{\mu\nu} - \frac{1}{2}Rg_{\mu\nu} + \Lambda g_{\mu\nu} = \frac{8\pi G}{c^4} T_{\mu\nu}
\end{equation}

\noindent where the term $T_{\mu\nu}$ on the RHS contains information about mass and energy, while everything on the LHS contains information about space and time - or rather spacetime. The equation mathematically proves Wheeler's statement. We see that Newtonian and Einsteinian ideas of fundamental entities have very little in common as the very meaning of terms change. Moreover, even the mathematical structure of these theories differ because Einstein imposes a non-Euclidean structure on spacetime, as opposed to Newton. We will discuss more examples of this across all of physics, when we consider physicists' arguments against incommensurability.

\section{What do Physicists have to say about Incommensurability?}

\noindent In 2002, the famous Russian trio of physicists - George Gamow, Dmitri Ivanenko and Lev Landau - published a paper titled \emph{World Constants and Limiting Transitions} \cite{article}. The paper talks about three things - \emph{theories} of physics (classical mechanics, quantum mechanics, etc.), \emph{world constants} (also called the fundamental constants of nature) and \emph{limiting transitions} (which is the idea that under different mathematical limits of these constants, one can make transitions from one theory to another).\\

\noindent Over the years, this idea has developed into a rich field of research which concerns itself with the unification of different physical theories into one all-encompassing framework called the theory of everything \cite{Manero2019-MANIOT-6}. The notion of world constants and limiting transitions has morphed into the pictorial representation called the cube of physics \cite{Padmanabhan2015}, which is depicted in figure 1. The theories in this model include classical mechanics (CM), quantum mechanics (QM), special relativity (SR), general relativity (GR), quantum field theory (QFT), Newtonian gravity (NG), non-relativistic quantum gravity (NRQG), and the theory of everything (TOE). These make up the vertices of the cube of physics. By themselves, these theories are just disjointed points having no correlation with each other. However, the effect of adding fundamental constants of nature to this model (i.e, the speed of light $c$, Planck's constant $\hbar$ and the universal gravitational constant $G$) is that of connecting these points. \\

\noindent In other words, we have theories, and we have bridges between these theories that are defined by the fundamental constants of nature. The way to think about this is as follows. Assume we have a consistent TOE which has all the constants ($c,\hbar,G$) present in its framework. Then in the absence of $\hbar$, the TOE reduces to GR, while in the absence of $c$, the TOE reduces to NRQG, and so on and so forth until we arrive at CM. If we run this process backwards, then we have a consistent framework to deduce a theory of everything from the already existing theories, i.e, one can unify all existing theories into a single theory. 

\begin{figure}[!h]%
\centering
\includegraphics[scale=0.8]{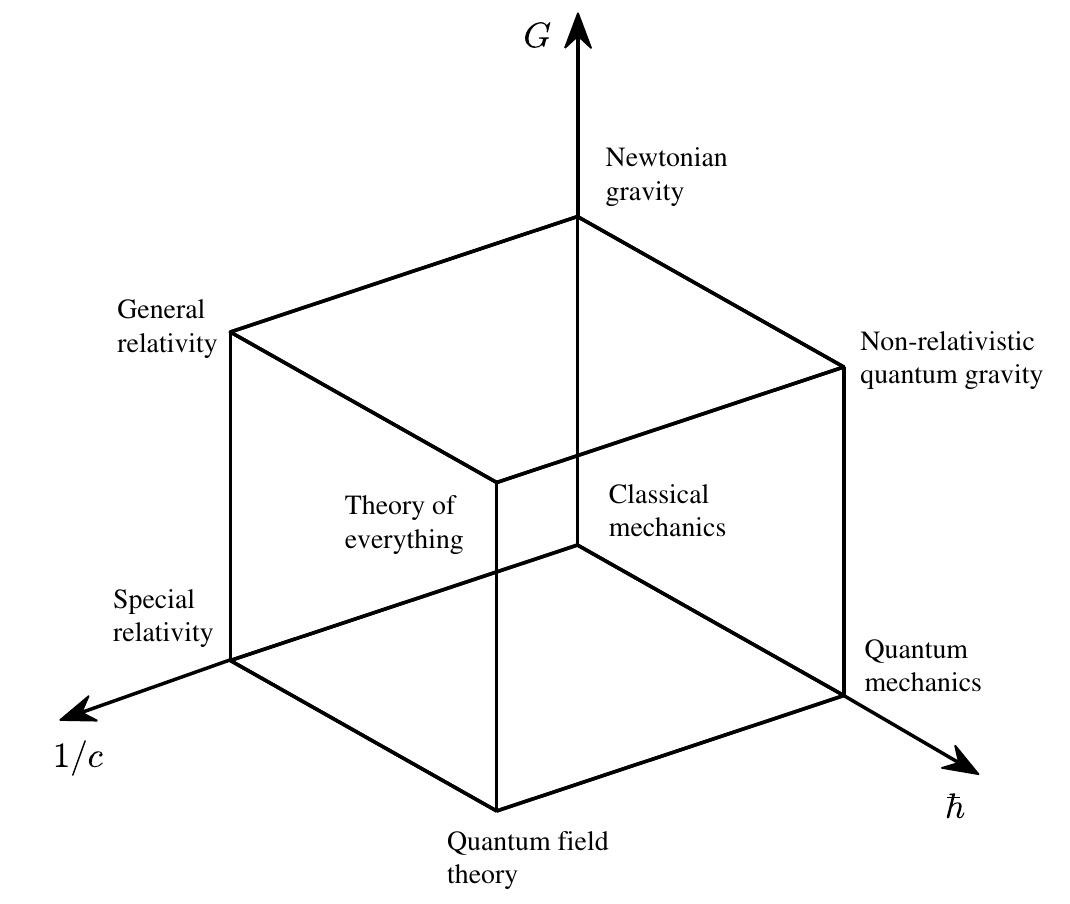}
\caption{The cube of physics describes physical theories and the transitions between these theories in terms of different conditions imposed on the fundamental constants of nature.}\label{fig1}
\end{figure}

\noindent The fact that the unification problem is probably pursued the most by physicists has something to say about the faith that physicists have on the cube of physics model. Physicists have shown that these transitions by \emph{the running of the constants} (as physicists like to call it) are true for already existing theories. For instance, SR does reduce to CM when $c \rightarrow \infty$. Hence, they inductively infer that the cube of physics is reliable, and that it will work for NRQG and TOE as well. What does this say about incommensurability? It seems as if it is possible to deductively infer one theory from another, unlike what Feyerabend said. At the outset, we get the idea that incommensurability may not be a problem after all. But after having a closer look, one can immediately find the flaws with this argument.\\

\noindent First of all, the conditions imposed on the fundamental constants are purely mathematical in nature, and carry no physical meaning. Consider the transition from quantum mechanics to classical mechanics. This is mathematically equivalent to the equation,

\begin{equation}\lim_{\hbar \to 0} \text{QM} = \text{CM}\end{equation}

\noindent This equation will only make sense to someone who does not care about the fundamental constants of nature. If one accepts that the Planck's constant $\hbar$ is a fundamental constant of nature, then the expression $\hbar \to 0$ makes no sense, because that would mean that the fundamental constant is no longer a constant, because its value changes! In this case, we conclude that such a limiting transition from quantum mechanics to classical mechanics is not possible. Hence, quantum mechanics and classical mechanics are incommensurable.\\

\noindent Secondly, even if we accept that $\hbar$ is not a fundamental constant of nature, and that its value could change, this does not prove anything about the truth of equation (3.1). It only implies that the equations of quantum mechanics (like Schrodinger's wave equation) reduce to the equations of classical mechanics (like Hamilton's equation). However, such mathematical deductions do not imply meaning invariance. The meanings of energy, mass, space, etc. in QM do not make a transition to their counterparts in CM in the limit where Planck's constant vanishes. So, even if $\hbar$ is not a constant, QM and CM are semantically incommensurable. In fact, this is exactly what physicists debate about in many conferences and lectures. In an ordinary undergraduate quantum mechanics lecture, no physicist will dare say that the concepts of QM have the same meaning as that of CM. On the other hand, they will (without hesitation) try to derive a mathematical relation between $[\hat{x},\hat{H}]$ (the quantum mechanical concept of motion) and \{$x,H$\} (the classical concept of motion).\\

\noindent These arguments will apply to any other transition described by the cube of physics, because they are all essentially of the same nature. But so far, we have only considered the physicists' case against semantic incommensurability. There are some popular arguments by physicists against methodological incommensurability as well. The most common one asserts the consideration of computational simplicity as a common measure to compare theories. Almost any classical physicist will say that the Hamiltonian and Lagrangian generalizations of mechanics are computationally much more simpler than the Newtonian counterpart, i.e, it is easier to solve problems using the action principle, rather than the force diagram approach of Newtonian mechanics. This is certainly true. In fact, the action principle has become a sort of consistency check for theoretical physics. Physicists say that if one cannot derive a theory from an action principle, then there is something wrong with the theory.\\

\noindent Again at the outset, this arguments discards methodological incommensurability. However, there is a flaw with this argument when one considers the subject of this argument. These arguments do not refer to different theories themselves, but to different mathematical formulations of the same theory. For instance, these arguments do not say that \emph{Newtonian mechanics} is better than \emph{general relativity} because of the computational simplicity of the former, but that the \emph{Lagrangian formulation} of classical mechanics (in the mathematical arena of configuration space) is better than the \emph{Newtonian formulation} (in the mathematical arena of physical space)\footnote{With quantum mechanics, the case is completely different, because its formulation in physical space (Bohmian mechanics) is altogether non-local.}. So, there is still no common base for comparison of theories. Methodological incommensurability persists.\\

\noindent Physicists often create a barrier of tests for new theories. This barrier includes concepts such as locality, Lorentz invariance, gauge invariance, etc. These are all certain mathematical tests for a theory. If the theory passes these tests, then it is a good theory. This might be used as an argument against methodological incommensurability, but the problem here is that this sounds exactly like the case with Newton's theory, where it became an exemplar for other theories. The new paradigms are tested against the laws of the existing paradigms, and thus there is no objective measure of comparison outside both these paradigms. Methodological incommensurability still persists.

\section{Concluding Remarks}

\noindent We have considered physicists' most powerful weapons against incommensurability and shown that these are either faulty weapons (i.e., they are inconsistent) or not weapons at all (i.e., they end up supporting the incommensurability thesis). These weapons are primarily used against methodological and semantic incommensurability. The validity of observational incommensurability is a bit unstable, but it can still be argued for in certain contexts.\\

\noindent The easiest way to argue for observational incommensurability is by comparing the notion of observations in classical and quantum physics. By definition, in classical physics, one is bound to observe classical objects (which have a certain minimum length scale or size). Likewise, in quantum physics, one is bound to observe subatomic particles or atoms (that are smaller than classical objects). This conveys the theory-ladenness of observations at the coarsest level. On the other hand, the very definition of a measurement or an observation is different in quantum mechanics \cite{RevModPhys.38.453,PhysRevD.2.2783}. An observation in quantum mechanics is supposed to permanently alter the state of the system that one is observing. This weirdness is not to be found in classical physics.\\

\noindent This might seem like a good argument for observational incommensurability, and one can always dig a little deeper by looking at more specific examples. But, as aforementioned, the only way to argue for observational incommensurability (at least to the best of my knowledge) is to prove it explicitly for every possible scenario. The answer to the question posed in the title of this paper is that even though methodological and semantic incommensurability can be considered to be existing \emph{globally} across all of physics, one is only allowed to argue for the \emph{local} existence of observational incommensurability within a few physical models, unless its existence is proved across all possible models.

\bibliographystyle{IEEEtran}
\bibliography{IEEEabrv,citations}
\nocite{*}


\end{document}